# Room-temperature dislocation plasticity in ceramics: Methods, Materials, and Mechanisms

Alexander Frisch[1]*, Chukwudalu Okafor[1], Oliver Preuß[2], Jiawen Zhang[3], Katsuyuki Matsunaga[4], Atsutomo Nakamura[5], Wenjun Lu[3], Xufei Fang[1,5]*

[1]Institute for Applied Materials, Karlsruhe Institute of Technology, Karlsruhe, Germany

[2]Department of Materials and Earth Sciences, Technical University of Darmstadt, Darmstadt, Germany

[3]Department of Mechanical and Energy Engineering, Southern University of Science and Technology, Shenzhen, P.R. China

[4]Department of Materials Physics, Nagoya University, Nagoya, Japan

[5]Department of Mechanical Science and Bioengineering, Graduate School of Engineering Science, Osaka University, Osaka, Japan

*Corresponding authors: alexander.frisch@kit.edu (AF); xufei.fang@kit.edu (XF)

**Abstract**

Dislocation-mediated plastic deformation in ceramic materials has sparked renewed research interest due to the technological potential of dislocations. Despite the long research history of dislocations as one-dimensional lattice defects in crystalline solids, the understanding of plastically deformable ceramics at room temperature seems lacking. The conventional view holds that ceramics are brittle, difficult to deform at room temperature and exhibit no dislocation plasticity except in small-scale testing such as nanoindentation and nano-/micropillar compression. In this review, we attempt to gather the evidence and reports of room-temperature dislocation plasticity in ceramics beyond the nano-/microscale, with a focus on meso-/macroscale plasticity. First, we present a mechanical deformation toolbox covering various experimental approaches for assessing the dislocation plasticity, with a focus on bulk plasticity. Second, we provide a materials toolbox listing 44 ceramic compounds that have been reported to exhibit dislocation plasticity at meso-/macroscale under ambient conditions. Finally, we discuss the mechanics of dislocations in ceramics, aiming to establish a foundation for predicting and discovering additional ceramics capable of room-temperature plastic deformation, thereby advancing the development of prospective dislocation-based technologies.

## 1. Introduction

Dislocations are line defects in crystalline solids and serve as the main carriers of plastic deformation. While dislocations are most widely known in metallic materials, early studies on their mechanics in ionic crystals such as LiF [1] significantly advanced the fundamental understanding of this material defect. In recent years, the renewed research activities in *dislocations in ceramics* have been inspired by the technological potential held by such one-dimensional line defects, namely "dislocation technology" [2-4], for the next-generation functional ceramics. For instance, dislocations interacting with ferroelectric domain walls can be utilized to modify the piezoelectric properties of ferroelectric materials [5]. Similarly, in electroceramics, increased electrical conductivity after plastic deformation has been attributed to the charged characteristics of dislocations in ionic crystals [6]. Furthermore, in plastically deformable semiconductors, the interaction between dislocations and photo-excited charge carriers enables the tuning of the material's bandgap through deformation [7, 8]. Most recently, localized magnetic order has been observed around dislocations in the ceramic $SrTiO_3$ [9]. These findings highlight the intricate interplay between mechanics and functionalities of ceramics, demonstrating the potential of dislocation engineering to create advanced ceramic materials with enhanced functionalities [4, 10].

These emerging research activities suggest that the role of dislocations in ceramics may have been much underappreciated. In general, ceramics are as considered brittle materials with limited plasticity, making them prone to stochastic and catastrophic failure. This typically low degree of plasticity has been attributed to several factors, including strong covalent/ionic bonding [11], larger Burgers vector compared to metals [12], limited slip systems, and grain boundary acting as effective barriers for slip transmission [13], and a low density of mobile dislocations [11]. The lack of room-temperature plasticity remains a significant bottleneck for the development of dislocation technology in ceramics, as dislocations do not nucleate and move as easily in most ceramics as in metals. To address this challenge, recent efforts by the authors have focused on developing the "deformation toolbox", i.e., achieving efficient dislocation engineering in ceramics at room temperature via mechanical deformation by focusing on circumventing dislocation nucleation while promoting dislocation multiplication in ceramics that exhibit good dislocation mobility, as in the case of using *mechanically seeded dislocations* [14]. This approach has proven effective in increasing the dislocation density and plastic deformation zone, although it has mainly been applied to model systems such as $SrTiO_3$ and MgO [15-18], which demonstrate excellent bulk plasticity at ambient conditions. In parallel, efforts have shifted towards developing a "materials toolbox" to identify additional ceramics capable of room-temperature plastic deformation, with $KTaO_3$ as a most recent example [19].

To lay the groundwork for the materials toolbox, here we aim to review the less-known or overlooked ceramic materials that are plastically deformable at room temperature, extending the past efforts by Sprackling [20] and Haasen [21, 22]. It should be noted that we did not include earlier works on slip in minerals from the 1920s-1930s [23-25], which were conducted even before the conceptualization of dislocations by Taylor [26], Orowan [27], and Polanyi [28] independently in 1934. It was observed that most reports and knowledge about plastically deformable ceramics at room temperature are scattered across various research fields and may have largely been forgotten. To this end, this article first presents the methods for mechanically introducing dislocations across the length scale but with a focus on meso/macroscale testing (**Section 2**), then categorizes the identified 44 ceramic compounds based on their crystal structure (**Section 3**), followed by analyzing the dislocation mechanisms **(Section 4)** with regard to their appearance in the discussed crystal structures. Finally, we discuss some strategies for predicting more plastically deformable ceramic compounds.

**Figure 1** illustrates the scope and the plastic deformation mechanisms in this work. As our focus here is on the plastic deformation of ceramics under ambient conditions, we do not cover studies using high-pressure torsion [29] or high confining pressure [30]. The material class of plastically deformable layered 2D Van-der-Waals chalcogenide crystals [31] shall be excluded from this review, as the main deformation mechanism of shearing weakly bonded lattice planes differs from dislocation-based plasticity. Similarly, nanocrystalline ceramics that plastically deform via grain-boundary sliding [32] are also not included. Additionally, plastic deformation mechanisms such as microcracking [33], phase transformation [34], and twinning [35] have been properly discussed in literature and will not be part of this review. Nor are ceramics exhibiting dislocation activity only subjected to nanomechanical testing such as in nanoindentation and nano-/micropillar compression, unless the small-scale dislocation plasticity can be scaled up and compared with the macroscale behavior [14, 36]. It is worth noting that the lesser likelihood of encountering defects in small deformation volumes and the much higher applied shear stresses at smaller scales can significantly facilitate dislocation plasticity, even in the case of diamond [37], which will be briefly discussed later.

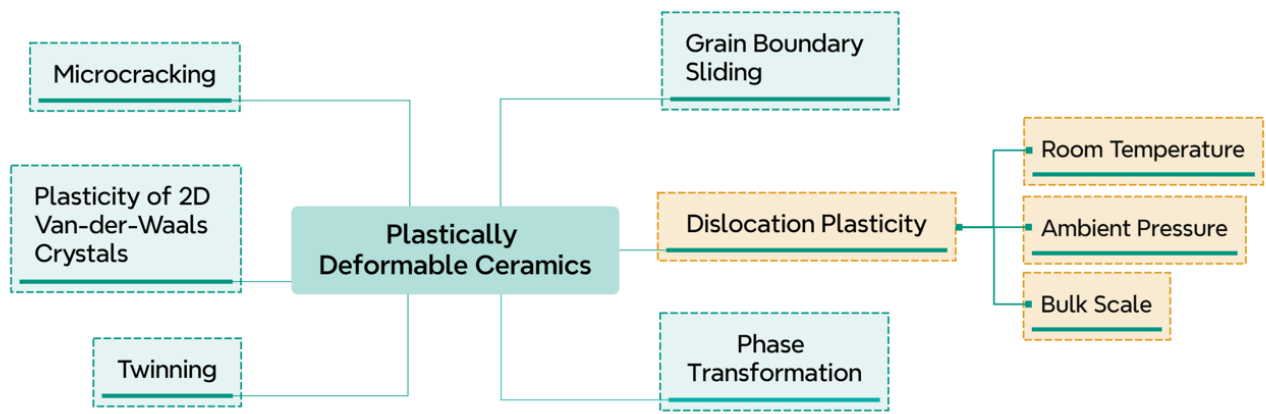

**Figure 1**. Overview of mechanisms contributing to plastic deformation in ceramic materials. The scope of this review, dislocation plasticity at room temperature, ambient pressure, and bulk scale, is highlighted in orange.

## 2. Methods

The methods for introducing dislocations into ceramics can generally be categorized into two main approaches: the processing route and the mechanical deformation route. The processing route includes techniques such as high-pressure high-temperature sintering [38, 39], flash sintering [40, 41], bicrystal bonding [42, 43], thin film growth [44], high-pressure torsion [45], or irradiation [46-48]. Here, we focus on the mechanical deformation approach for ceramics. **Figure 2** highlights the key features of dislocation-mediated plastic deformation introduced by the most common deformation methods: (A) nanoindentation; (B) Vickers indentation; (C) cyclic Brinell indentation; (D) cyclic Brinell scratching; and (E) uniaxial bulk compression. These methods are arranged by the representative plastic zone size, increasing in the length scale. Other techniques such as surface grinding and polishing, sandblasting and shot peening, and bending tests will be briefly discussed later in this section.

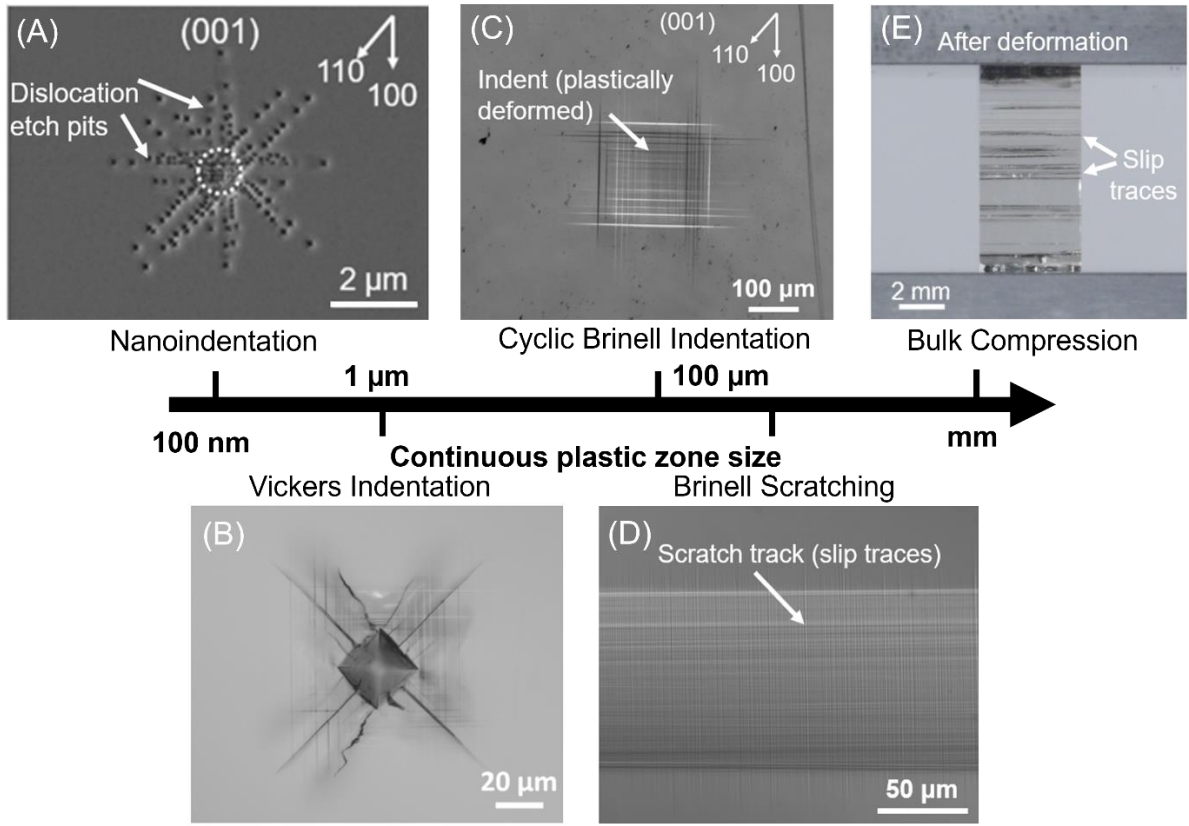

**Figure 2**. Methods for mechanically imprinting dislocations through deformation sorted by increasing plastic zone size. Single-crystal SrTiO$_3$ is used as a model material for demonstration: (A) Nanoindentation. The dislocations are visualized by the etch pits. Image reprinted and modified with permission from Ref. [49]. (B) Vickers indentation. Slip traces as well as cracks are visible around the indent imprint. (C) Cyclic Brinell indentation. The dense pattern of slip traces indicates the presence of a high dislocation density. (D) Cyclic Brinell scratching. Only a fraction of the track is pictured as its length was in the mm range. (E) Uniaxial bulk compression. Slip band formation is visualized by the dark lines. Images (C-E) are reprinted and modified with permission from Ref. [10].

### 2.1. Nanoindentation

For completeness, the nanoindentation method is included in **Fig. 2**, although the small-scale plasticity is not the primary focus of this work. Initially developed to measure the hardness and modulus of various materials [50], nanoindentation has proven to be a simple but powerful technique for probing dislocation mechanisms, such as dislocation nucleation [51] and motion [52, 53]. This is achieved by carefully evaluating the load-displacement curves recorded in the machine on indenting a well-polished flat surface. Due to the very local deformation volume being probed (hence a high chance of testing a flaw-free region) using a sharp indenter (e.g., diamond tip with a tip radius in the range of hundreds of nanometers), the induced local shear stress can peak above the theoretical shear strength to homogeneously nucleate dislocations [51]. Extensive experimental evidence and discussions are available on using this method to study the localized dislocation plasticity on a wide range of ceramics at room temperature.

However, due to the extremely high local shear stress underneath the sharp indenters, many ceramics that exhibit plastic deformation in nanoindentation tests do not demonstrate bulk plasticity. With the length scale increasing, the majority of ceramic samples fracture in a brittle manner due to insufficient shear stress to mobilize the dislocations prior to the onset of crack initiation and propagation. A representative experimental study was conducted on $TiO_2$ and $Al_2O_3$, both of which display nanoindentation plasticity at room temperature, but crack immediately when tested at larger scales [51]. Therefore, nanoindentation, being a local probing technique, is not a suitable tool for direct experimental evaluation of the macroscale plastic behavior of ceramics. Nevertheless, nanoindentation can provide valuable insights into the fundamental dislocation mechanisms, which are critical to be linked to the understanding of bulk plasticity in ceramics [54]. For instance, as illustrated in **Fig. 2A**, using the spacing between the dislocation etch pits, it is possible to estimate in these materials the lattice friction stress, namely the resistance to the motion of dislocations, which is a critical parameter for assessing the capability of the bulk deformability of such materials [14], as in the case of MgO and $SrTiO_3$ [52, 53].

**2.2. Vickers indentation**

Vickers indentation, widely used for hardness measurement, involves pressing a pyramidal-shaped diamond indenter into the material's surface. This cost-effective method can also be utilized for investigating dislocation activity in ceramics at room temperature. The sharp indenter tip and edges induce locally high stress concentrations underneath the indenter, which facilitate dislocation multiplication and motion in the plastically deformed volume, resulting in slip traces that can be viewed by optical microscopy [55]. However, accompanying crack formation including radial/median/lateral cracks is commonly seen depending on the materials, tip geometries, and loading conditions [56, 57]. For instance, **Fig. 2B** illustrates a Vickers indent on a polished (001) $SrTiO_3$ sample, with clear evidence of dislocation slip traces but also cracks. This method provides a quick assessment of the room-temperature macroscopic deformability of a well-polished ceramic: the plastically deformable ceramics will produce slip traces around the indent. These slip traces serve as fingerprints of the room-temperature active slip systems. In contrast, brittle ceramics typically display crack formation without visible slip traces.

## 2.3. Brinell indentation

Brinell indentation, which uses a spherical indenter with a tip radius of several millimeters, is an alternative to Vickers indentation for introducing dislocations without initiating cracks. This method has been used to study dislocation behavior in ceramics like LiF [58], MgO [59], $CaF_2$ [60], and $SrTiO_3$ [16]. The hydrostatic compressive stress components under the indenter tip help suppress crack formation, while the shear stress components promote dislocation multiplication and motion. This approach has another advantage of tuning the dislocation density by adjusting the number of indentation cycles. Repeatedly indenting the same region increases dislocation density, reaching values as high as ~$10^{13}$ m$^{-2}$ [16]. **Fig. 2C** illustrates a plastic zone created by cyclic Brinell indentation with increased dislocation density [10]. Depending on the tip radius of the Brinell indenter, this technique can generate a dislocation-rich plastic zone up to a few hundreds of micrometers in all dimensions [16].

## 2.4 Brinell scratching

An extension of Brinell indentation is the Brinell scratching technique, where the spherical indenter is repeatedly dragged over the sample surface under applied load. This approach produces a larger plastic zone, with dislocation density increasing with the number of scratching passes [17]. For instance, **Fig. 2D** presents slip traces from a scratch track after ten passes. Dislocation densities of up to ~$10^{15}$ m$^{-2}$ and plastic zones several hundred micrometers deep can be achieved [18]. By overlapping the scratch tracks, this method can easily cover an entire sample surface with dislocations penetrating hundreds of micrometers into the sample [17].

Although arguably not as well established as uniaxial compression or Vickers indentation, cyclic Brinell indentation and scratching have a long history in the deformation of ceramic crystals. Both techniques were reported to deform single-crystal LiF, MgO, and even $SrTiO_3$ in the 1970s and 80s by Brookes and Shaw [59, 61-64], although the focus did not lie much on tuning the dislocation density or the plastic zone size for any dislocation functionality evaluation. More importantly, cyclic Brinell indentation and scratching have been successfully applied to introduce dislocations into polycrystalline $SrTiO_3$, $Mn_xZn_yFe_{3-x-y}O_4$, and $Ni_xZn_{1-x}Fe_2O_4$ samples without crack formation [65, 66], suggesting these two methods, in combination, may hold much greater application potential. It was demonstrated recently that by rolling instead of scratching a sphere across a ceramic surface, dislocations can also be introduced into ceramics, resulting in comparable wear tracks [67].

## 2.5. Uniaxial bulk compression

A straightforward demonstration for room-temperature plasticity of a given ceramic material is by bulk test. As most ceramic materials fail catastrophically under tensile loading, primarily due to crack propagation from flaws under tension [68], bulk compression tests are usually chosen to avoid rapid fracture. It is therefore not surprising that the earliest experiments conducted on examining the dislocation plasticity in ceramics were by uniaxial bulk compression [69, 70]. In this technique a sample is compressed by moving a spindle-driven crosshead, usually with a constant strain rate of ~$10^{-4}$ $s^{-1}$ or lower. A single-crystal $SrTiO_3$ sample tested in such a device is demonstrated in **Fig. 2E** [10]. For instance, all the three perovskite oxides that are capable of room-temperature bulk plasticity, namely, $SrTiO_3$ [71], $KNbO_3$ [72], and $KTaO_3$ [19], have been verified by uniaxial bulk compression. With this method, the stress-strain curves can be obtained to extract important material parameters such as the Young's modulus and the yield strength. Furthermore, by changing the strain rates e.g. from $10^{-5}$ $s^{-1}$ to $10^{-1}$ $s^{-1}$ [73], the different dislocation mechanisms activated during the plastic deformation can be investigated.

While obtaining a large plastic volume and determining the bulk yield strength via this technique makes it very appealing to investigate the ceramics' bulk plastic deformation, the shortcomings of this method are also evident. First of all, most crystals that are suitable for bulk compression are very expensive or difficult to grow. Second, while seemingly simple, the method can be challenging to master while testing the crystals. One of the most critical aspects is the good alignment between the loading cell and the sample (with parallel top/bottom surface) to avoid stress concentration at the sample top/bottom edges, where very often the cracks originate [19]. Another aspect is that these crystals often deform non-uniformly under uniaxial bulk compression, producing discrete and confined slip bands [69, 73, 74] that are adjacent to regions that are virtually undeformed, as can be seen in **Fig. 2E**. While polycrystalline samples are much cheaper and easy to fabricate for bulk compression tests, the limited slip systems in most ceramics at room temperature do not fulfill the von Mises or Taylor criteria of five independent active slip systems [75-77], hence cracks are easily formed at the grain boundaries to induce fracture of the samples. These critical factors make bulk compression of single-crystal ceramic samples a rather time-consuming and cost-intensive endeavor, especially for systematic and statistical investigations at room temperature.

**2.6 Other deformation methods**

In addition to nanoindentation method discussed above, a few more near-surface techniques for dislocation introduction shall be mentioned here. Those techniques are distinct from the previous methods, as they only introduce dislocations into the first few micrometers of the near-surface regions of the samples. For instance, surface grinding/polishing was used to introduce high-density dislocations on ceramic surfaces, in the range of $10^{15}$ $m^{-2}$ [15, 78, 79]. The micro-sized particles on the grinding papers induce local stress concentration to generate dislocations [14], with a steep gradient in the stress field, hence the dislocation density gradient along the depth up to a few micrometers [10]. Another technique for near-surface dislocation introduction over large sample areas is sand blasting or shot peening. Here, a ceramic's surface is blasted with accelerated hard particles, resulting in large compressive stresses at several GPa to enhance the surface strength of ceramics. Nevertheless, the role of the dislocations introduced using this method was poorly discussed or neglected in the past. Again, the dislocation density drops rapidly within the first ~100 μm from the surface [80, 81].

The above list of deformation methods is not exhaustive as the less popular methods such as bending tests [82] has not been discussed so far. This method requires single crystals of relatively large size and tensile stress on one side during the bending tests. As the flaws and cracks in ceramic materials are prone to open under tensile stress, this method requires careful handling of the sample and testing, as cracks may easily propagate to fracture the bending bars. Nevertheless, bending tests can form periodic, discrete slip bands on the sample surface, which was most recently used for probing the dislocation-induced local magnetic order in $SrTiO_3$ [9].

The ultimate validation for *ductile* ceramics in bulk at room temperature will be the tensile testing. For oxides, it appears to be so far only achieved in bulk single-crystal MgO by Stokes et al. [83], in bulk single-crystal LiF by Majumdar et al. [84], and in bulk polycrystalline AgCl by Lloyd et al. [85]. To this end, in the case of MgO, the samples were subjected to sprinkling (to generate a lot of dislocations as sources) before tensile loading. As in the case of LiF, compression test was first conducted (again, to induce dislocations as sources) before tensile loading was applied, where cyclic compression/tension was also achievable. For polycrystalline AgCl, no further preparation was necessary to achieve bulk-scale ductility in tensile testing. This is, however, due to the exceptional amount of active slip systems in this compound, which will be explained in the following **Section 3**. Nevertheless, besides overcoming the dislocation source issue by pre-deformation, it is critical to avoid stress concentration at the sample/holder junction during tensile testing, which can be a major challenge for sample fabrication and clamping.

As an overview, the various mechanical deformation methods and their advantages and limitations are briefly summarized in **Table 1**.

**Table 1.** Different mechanical deformation methods for dislocation engineering in ceramics.

| Method | Advantages | Disadvantages |
|---|---|---|
| **Nanoindentation** | <ul><li>Load-displacement curve</li><li>All microstructures</li><li>Cracking avoidable</li><li>High dislocation density</li></ul> | <ul><li>Localized</li><li>Size effects</li></ul> |
| **Vickers indentation** | <ul><li>All microstructures</li><li>Simple & fast to test</li></ul> | <ul><li>Cracking around indent</li><li>Localized</li><li>Complex stress field</li><li>No stress-strain curves</li></ul> |
| **Cyclic Brinell indentation** | <ul><li>All microstructures</li><li>Simple & fast to test</li><li>Controllable dislocation density</li><li>Cracking avoidable</li></ul> | <ul><li>Complex stress field</li><li>Localized</li></ul> |
| **Cyclic Brinell scratching** | <ul><li>All microstructures</li><li>Cracking avoidable</li></ul> | <ul><li>Complex stress field</li><li>Localized</li></ul> |

|  | • Controllable dislocation density<br>• High dislocation density |  |
|---|---|---|
| **Uniaxial compression** | • Large deformed volume<br>• Stress-strain curve available | • Discrete slip bands<br>• Cracking from edges<br>• Time consuming |
| **Uniaxial tension** | • Large deformed volume<br>• Stress-strain curve available | • Discrete slip band<br>• Easy to fracture, cracking from sample clamping<br>• Time consuming<br>• Specimen fabrication difficult |
| **Surface grinding/ polishing/ sand blasting/ shot peening** | • All microstructures<br>• Easy & fast | • Uncontrollable<br>• Limited plastic zone in the sample skin area |
| **Bending test** | • Deformation data | • Very localized<br>• Size effects |

## 3. Materials

In this section, we summarize a list of 44 ceramic compounds that have been reported to exhibit dislocation plasticity at ambient conditions in the millimeter range. It is important to note that for all of these compounds, the dislocation-based plastic deformation was reported on single-crystal samples, as most polycrystalline ceramic samples suffer from the limited independent slip systems at room temperature [4, 13, 65] as mentioned above.

For simplicity, with respect to identification of their slip systems, they are grouped by crystal structure. Among the following seven crystal structures, the group of rock-salt structure crystals is by far the largest. This group is comprised mostly of alkali halides, simple metal oxides, and a few chalcogenide semiconductors. Then it follows the group of perovskite oxides [10, 86], which are of greater interest due to their functional properties. Next up are several semiconductors with sphalerite and wurtzite structures, whose deformation has been investigated mostly due to the photoplastic and electroplastic effects as well as the influence of dislocations on the band structure [7, 87-89]. A smaller group is the fluorite structure, with $CaF_2$ and $BaF_2$. Their deformation were investigated in the past due to their structural similarity to $ZrO_2$ and $UO_2$, both of which, in contrast, do not display room-temperature bulk activity [90, 91]. Another small group contains compounds of the cesium chloride structure, which have not been investigated much recently but exhibit excellent deformability with critical resolved shear stresses under 1 MPa [92, 93]. Lastly, there are some deformable magnetic oxides with spinel structure, whose deformation has not been investigated since 1990s, and up until then only rarely [94, 95]. As 44 compounds are too many to go into the details, only a few prominent examples are discussed later. Here some of the dislocation mechanisms will already be briefly mentioned and more detailed explanations will follow in **Section 4**.

### 3.1 Rock salt structure

**Table 2.** Rock salt structure ceramic compounds that exhibit dislocation-mediated plastic deformation at ambient conditions at bulk scale. The crystal structure is schematically depicted with a representative material. The active slip systems are summarized based on the literature. The same applies for **Tables 3-8**.

| Rock Salt Structure | Compound | Active Slip Systems | References |
|---|---|---|---|
| | **LiF** | $\{110\}\langle110\rangle$ | **[74, 96-98]** |
| | **LiCl*** | $\{110\}\langle110\rangle$ | **[99]** |
| | **LiBr*** | $\{110\}\langle110\rangle$ | **[99]** |
| | **NaF** | $\{110\}\langle110\rangle$ | **[98, 100]** |
| | **NaCl** | $\{110\}\langle110\rangle$ | **[98, 101]** |
| | **NaBr** | $\{110\}\langle110\rangle$ | **[98]** |
| | **NaI** | $\{110\}\langle110\rangle$ | **[98, 101, 102]** |
| | **KCl** | $\{110\}\langle110\rangle$ | **[96, 98, 101, 103]** |
| | **KBr** | $\{110\}\langle110\rangle$ | **[96, 98, 103]** |
| | **KI** | $\{110\}\langle110\rangle$ | **[96, 98, 101, 104]** |
| | **RbCl** | $\{110\}\langle110\rangle$ | **[105]** |
| | **RbI** | $\{110\}\langle110\rangle$ | **[104]** |
| | **MgO** | $\{110\}\langle110\rangle$ | **[17, 98, 103, 106, 107]** |
| | **CaO** | $\{110\}\langle110\rangle$ | **[98, 106, 108]** |
| | **CoO** | $\{110\}\langle110\rangle$ | **[109, 110]** |
| | **NiO** | $\{110\}\langle110\rangle$ | **[110-112]** |
| | **SrO** | $\{110\}\langle110\rangle$ | **[106]** |
| | **AgCl** | $\{001\}\langle110\rangle$ $\{110\}\langle110\rangle$ $\{111\}\langle110\rangle$ | **[113-115]** |
| | **AgBr** | $\{110\}\langle110\rangle$ | **[114, 116]** |
| | **SmS** | $\{110\}\langle110\rangle$ | **[117, 118]** |
| | **PbS** | $\{100\}\langle110\rangle$ | **[96, 119]** |
| | **PbTe** | $\{100\}\langle110\rangle$ | **[96]** |

*Note: for these materials, no bulk compression data was found but bending test results.*

Most of the earliest works on dislocations in ceramic materials were conducted on materials of the rock salt structure [97, 99, 100, 120]. This structure can be described as two interpenetrating face-centered cubic lattices of oppositely charged ions (see illustration in **Table 2**). The prevalent slip systems for this group of materials are of the {110}<110> type [98, 121, 122], which corresponds to six physically distinct slip systems, from which only two are independent [77]. AgCl and AgBr are exceptions by having dislocations also gliding in the {001}<110> and {111}<110> glide system [115]. Some of the earliest investigations in the dislocation mechanisms were conducted in the 1950-60s by Gilman and Johnston with their classic dislocation etch pits studies on LiF. They were the first to observe isolated moving dislocations as well as dislocation multiplication, and much of today's understanding of dislocation motion and multiplication in ceramics comes from these seminal works [74, 96, 97, 123]. Almost all of the rock salt structure crystals on this list were demonstrated to deform in uniaxial bulk compression. The exceptions on this list are LiCl and LiBr, which displayed dislocation activity after bending experiments [99].

Studied for its deformation behavior even before the term *"dislocation"* was coined [124], NaCl has been just as important in dislocation research in the past, especially for the electrostatic properties of dislocations in ionic crystals, as first discussed by Eshelby et al. in 1958 [125] and later comprehensively reviewed by Whitworth in 1975 [126]. It was demonstrated back then, that the dislocations of the {110}<110> slip systems in the rock salt structure are inherently charge-neutral but could acquire charges by interacting with point defects [126]. Furthermore, in 1976, Sprackling summarized the plastically deformable rock-salt structure crystals and analyzed their deformation behavior [20], and a decade later, some of the first atomistic simulations on dislocation core structure and point defect interactions were conducted by Rabier and Puls [127, 128].

Due to its stability, ease of handling, and room-temperature bulk deformability, MgO has been an excellent model oxide for dislocation research, both experimentally and theoretically, as reviewed by Amodeo et al. [107]. Some of the earliest works on MgO can trace back to the 1960s, where Argon and Orowan investigated the slip band intersections in MgO after bulk compression [69]. Much later, using nanoindentation, Tromas et al. investigated the dislocation nucleation in MgO [129]. The understanding of dislocations in MgO has been further improved in recent years by computational simulations [121, 122, 130]. New insights about the exact core structures of edge and screw dislocations, the mechanism of dislocation motion, and the cross-slip behavior in MgO have been gained thanks to the recent simulation endeavor, contributing to the knowledge of dislocations in ceramics. Detailed discussions will follow in **Section 4**.

### 3.2 Perovskite structure

**Table 3.** Oxides with perovskite structure with room temperature bulk plasticity.

| Perovskite Structure | Compound | Active Slip Systems | References |
| --- | --- | --- | --- |
| | **SrTiO$_3$** | {110}⟨110⟩ | **[14, 71]** |
| | **KNbO$_3$** | {110}⟨110⟩ | **[72, 86, 131, 132]** |
| | **KTaO$_3$** | {110}⟨110⟩ | **[19, 133]** |

The current renewed research interest in dislocations in ceramics has been partly sparked by dislocations in perovskite oxides. These materials are well known for their versatile physical properties, with very recent studies on the dislocation influence on ionic conductivity [134], piezoelectric coefficient [5, 135], or local magnetic order [9]. SrTiO$_3$ is the first reported perovskite oxide (2001) that exhibits bulk compression plasticity at room temperature [71]. It has been used as a model system for a rich variety of dislocation investigations, ranging from the core structure [131, 136, 137], impact on mechanical properties [16, 18, 132], and physical properties as aforementioned. Still, there are new insights in the dislocations in this material up to date, as most recently addressed by Hirel et al. [138] on its dislocation charge character and mobility. Their simulation revealed the dominating charge-neutral nature of the mobile dislocations in SrTiO$_3$ at low temperature. Nevertheless, these dislocations may acquire charges by interacting with point defects. This is surprisingly analogous to the dislocation behavior in the alkali halides (e.g., NaCl) discussed above. Note, that both NaCl and SrTiO$_3$ have a cubic structure and the {110}<110> slip systems are active at room temperature.

The second room-temperature deformable perovskite, discovered in 2016, is KNbO$_3$ [72]. While its dislocations behave very similarly to those in SrTiO$_3$, KNbO$_3$ is orthorhombic or pseudo-cubic, and hence ferroelectric at room temperature. Its ferroelectric domain structure has been revealed to be influenced by the presence of dislocations [86, 132]. The most recent, i.e., the third room-temperature deformable perovskite, KTaO$_3$, has been discovered by the current authors [19] and confirmed in parallel by Khayr et al. [133]. The latter group investigated the physical property of the dislocation-induced local ferroelectric order. Like SrTiO$_3$, both KNbO$_3$ and KTaO$_3$ can be plastically deformed by uniaxial bulk compression [19, 72, 86].

### 3.3 Sphalerite structure

**Table 4.** Sphalerite structure ceramic compounds with room temperature bulk plasticity.

| **Sphalerite Structure** | **Compound** | **Active Slip Systems** | **References** |
|---|---|---|---|
| | **ZnS** | {111}⟨110⟩ | **[7, 139-141]** |
| | **ZnSe** | {111}⟨110⟩ | **[139, 141-143]** |
| | **ZnTe** | {111}⟨110⟩ | **[139, 141]** |

| | | |
|---|---|---|
| **CdTe** | {111}⟨110⟩ | **[139, 141]** |
| **CuCl** | {111}⟨110⟩ | **[144]** |
| **CuBr** | {111}⟨110⟩ | **[145]** |

The dislocation-mediated deformation of semiconductors of the sphalerite structure has gathered great attention in past literature and the present [140, 146-148]. The sphalerite structure can be seen as a face-centered cubic lattice of one atomic species with half of the tetrahedral voids filled with the other species, as depicted in **Table 4**. The active slip systems belong to the {111}<110> family, although the dislocations readily split into Shockley partials [7], a concept that will be explored more in the following section. The compounds zinc sulfide (ZnS) and zinc selenide (ZnSe) have been investigated since the 1970s for their photoplastic and electroplastic properties [88, 143, 146]. Most recent bulk compression tests by Oshima et al. [7] revealed that single-crystal ZnS can deform up to an astonishing ~45% strain in complete darkness before the crystal fractures, while the crystals fractures at ~6% strain under light. This behavior, coined as photoplasticity, is attributed to photo-excited carriers interacting with dislocations, and in turn the dislocations change the optical properties, as observed in the band gap change after deformation [7]. Furthermore, it was recently demonstrated that dislocations in ZnS become mobile by applying an electric field, opening a new way to facilitate dislocation motion apart from mechanical loading [87]. Apart from ZnS, the photo- and electroplastic properties of some of the sphalerite-type (as well as some wurtzite-type) semiconductors were reviewed by Osip'yan et al. in 1986 [89]. All of the listed sphalerite compounds deform plastically under uniaxial bulk compression [139, 144, 145].

### 3.4 Wurtzite structure

**Table 5.** Wurtzite structure ceramic compounds with room temperature bulk plasticity.

| **Wurtzite Structure** | **Compound** | **Active Slip Systems** | **References** |
|---|---|---|---|
| | **ZnO** | {1000}⟨1120⟩<br>{1100}⟨1120⟩ | **[139, 141]** |
| | **CdS** | {1000}⟨1120⟩<br>{1100}⟨1120⟩ | **[139, 141, 149]** |
| | **CdSe** | {1000}⟨1120⟩<br>{1100}⟨1120⟩ | **[139, 141]** |

Structurally similar to the cubic sphalerite lattice, the hexagonal wurtzite structure produces some room-temperature deformable semiconductors, that are listed in **Table 5**. In this structure, deformation was mostly reported to occur in the {1000}⟨1120⟩ slip system (basal slip) and in the {1100}⟨1120⟩ slip system (prismatic slip) [150]. Here these three deformable materials were also investigated for their photo- and electroplastic properties in the 1980s [89]. Among them is ZnO, an important wide bandgap semiconductor, has recently been studied using photoindentation (nanoindentation combined with light illumination) [151] for further investigation of the mechanical behavior of wurtzite structure materials [152]. Like the sphalerite compounds, the uniaxial bulk compressibility of the listed wurtzite compounds was confirmed [139].

### 3.5 Fluorite structure

**Table 6.** Fluorite structure ceramic compounds with room temperature bulk plasticity.

| Fluorite Structure | Compound | Active Slip Systems | References |
|---|---|---|---|
| | $CaF_2$ | {001}⟨110⟩ | **[17, 60, 153]** |
| | $BaF_2$ | {001}⟨110⟩ | **[154-156]** |

Of all the compounds crystallizing in the fluorite structure, only two have been reported to deform plastically under ambient conditions and uniaxial loading, $BaF_2$ and $CaF_2$, as depicted in **Table 6**. Dislocations in $CaF_2$ glide in the {001}<110> slip systems at room temperature [77] and more recently, the role of dislocations in micromachining $CaF_2$ has been investigated [157]. The lack of room-temperature plastically deformable members of this structure, however, does not entail a lack of understanding of dislocations in the fluorite structure. This is mainly due to the other prominent members $ZrO_2$ and $UO_2$, which do not deform plastically at room-temperature. Due to their application in ionic conduction ($ZnO_2$) and nuclear application ($UO_2$), the study of dislocations in fluorite structure has fueled insights into dislocations in $CaF_2$ as well [90, 91]. Furthermore, through recent simulation efforts, the dislocation core structure of $UO_2$, the mobility mechanisms and dislocation interactions have been revealed in this structure [158-160].

### 3.6 Cesium chloride structure

**Table 7.** Cesium chloride structure ceramic compounds with room temperature bulk plasticity.

| Cesium Chloride Structure | Compound | Active Slip Systems | References |
|---|---|---|---|
| | **CsBr** | $\{110\}\langle100\rangle$ | **[93, 161]** |
| | **CsI** | $\{110\}\langle100\rangle$ | **[92, 104, 161, 162]** |
| | **Tl(Cl,Br)*** | $\{110\}\langle100\rangle$ | **[163, 164]** |
| | **Tl(Br,I)*** | $\{110\}\langle100\rangle$ | **[163-165]** |

*For these materials, no bulk compression data was found.*

Dislocation studies in ionic compounds of the cesium chloride structure is not as abundant as the previous structures. The cesium chloride structure can be described as two interpenetrating cubic primitive lattices, as depicted in **Table 7**. Its room-temperature slip systems are {110}<100>. The listed compounds here have their uses in optics or scintillation [166, 167], but no changes in the physical properties with increasing deformation have been reported to the best of our knowledge. The dislocation-based mechanical behavior has been investigated in some detail between 1956 and 1985, with insights about the active slip system, core structure, and low-temperature glide mechanism investigated and discussed [93, 163]. CsBr and CsI have been shown to be deformable under uniaxial compression [93, 162]. No such studies are available for the Thallium-containing compounds, they were, however, indented with a needle, which produced macroscopically visible slip traces, indicating their room-temperature plastic deformability [164].

### 3.7 Spinel structure

**Table 8.** Spinel structure ceramic compounds with room temperature bulk plasticity.

| Spinel Structure | Compound | Active Slip Systems | References |
|---|---|---|---|
| | **$Fe_3O_4$*** | $\{001\}\langle110\rangle$ $\{111\}\langle110\rangle$ | **[168]** |
| | **$Mn_xZn_yFe_{3-x-y}O_4$*** | $\{110\}\langle110\rangle$ $\{111\}\langle110\rangle$ | **[66, 95, 169-171]** |
| | **$Ni_xFe_{3-x}O_4$*** | $\{111\}\langle110\rangle$ | **[94]** |
| | **$Ni_xZn_{1-x}Fe_2O_4$*** | $\{001\}\langle110\rangle$ $\{111\}\langle110\rangle$ | **[66]** |

*For these materials, no bulk compression data was found.*

Little is known about dislocations at room temperature in the spinel structure. Besides a few reports on the room-temperature deformability upon Vickers indentation on $Fe_3O_4$ and $Ni_xFe_{3-x}O_4$, Brinell indentation, Brinell scratching, or "sphere rolling" (similar to Brinell scratching, but the indenter tip can rotate) on $Mn_xZn_yFe_{3-x-y}O_4$ and $Ni_xZn_{1-x}Fe_2O_4$, the room-temperature dislocation behavior in these materials were not investigated in depth. Interestingly, slip systems of this structure have been reported to be {100}<110>, {110}<110>, or {111}<110>, with different authors claiming one or two of them to be active at room temperature [66, 94, 168]. All of the spinel compounds listed in **Table 8** are ferrimagnetic, with many studies on their magnetic properties [172, 173]. The properties of their dislocations, however, like their dislocation core structure or mechanisms of dislocation motion have not yet been studied in detail, leaving room for future studies on the plastic deformation of these materials.

For all of the 44 listed ceramic compounds, the dislocation-based plastic deformation has been reported on single crystals. Additional evidence of room-temperature plasticity in polycrystalline samples of AgCl [85], $SrTiO_3$ [65], $Mn_xZn_yFe_{3-x-y}O_4$ [66], and $Ni_xZn_{1-x}Fe_2O_4$ [66] have been reported. The most notable exception here is AgCl, of which polycrystalline sample were successfully deformed in tensile testing. The amount of active slip systems in this compound allows it to overcome the von Mises or Taylor criteria, allowing for bulk-scale polycrystalline deformability. This, however, is not applicable to most other ceramic compounds. Nevertheless, plastic deformation in the near surface region of some polycrystalline samples is achievable at room temperature. For instance, coarse-grained, polycrystalline $SrTiO_3$ was recently demonstrated to exhibit surface dislocation plasticity across various grain boundaries without forming cracks via cyclic Brinell indentation and scratching [65]. Similar results were reported from Brinell scratching on $Mn_xZn_yFe_{3-x-y}O_4$ [66], and $Ni_xZn_{1-x}Fe_2O_4$ [66]. This is achieved by deforming on the free surface, which relaxes the von Mises or Taylor criterion. It is expected, that such near-surface deformation techniques can be applied to the other ceramic compounds summarized in this section.

## 4. Mechanisms

As the previous sections on crystal structures demonstrate, there are significant differences among the deformable crystal structures, more so their slip systems and dislocation structures. This compilation is intriguing in the sense of paving the road for understanding the underlying mechanisms that govern room-temperature dislocation mechanisms and finding common ground to predict and identify more deformable ceramics. Previous efforts have been made to summarize the relevant mechanisms of dislocations in ceramics, however, the focus was predominantly on high-temperature deformation of polycrystalline ceramics [174, 175]. Here, in accordance with previous practices [4, 74], it has proven most useful to divide the mechanisms into dislocation nucleation, dislocation motion, and dislocation multiplication. The structure for this section is visualized in **Fig. 3**.

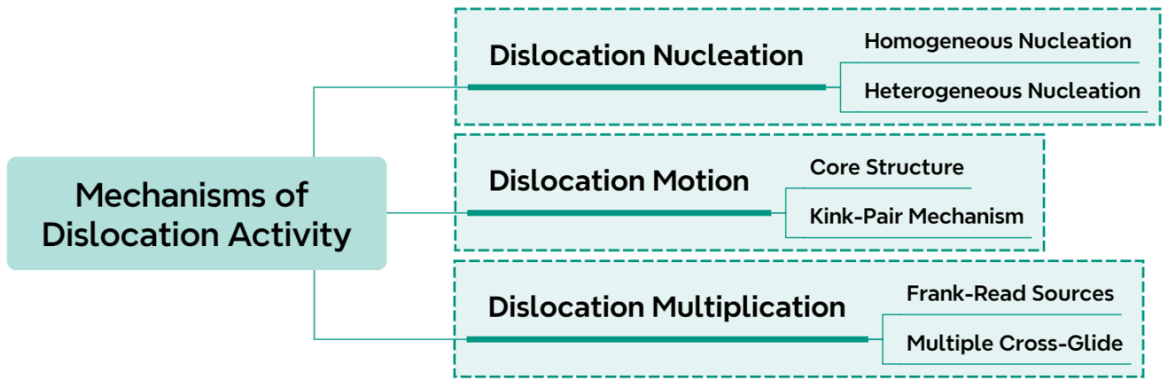

**Figure 3.** Overview over the mechanisms of dislocation activity

### 4.1 Dislocation Nucleation

Dislocation nucleation describes the introduction of dislocations into a volume previously dislocation-free. It concerns homogeneous nucleation and heterogeneous nucleation [13]. Homogeneous dislocation nucleation is the generation of a new dislocation from a defect-free crystal lattice. This process requires shear stresses approaching the theoretical shear strength (~$G/2\pi$, with $G$ being the shear modulus), and is usually only observed in nanomechanical testing [51, 176]. At bulk scale, stresses of this magnitude would not have been achieved before fracture or crack propagation from pre-existing flaws (e.g., microcracks, pores, and voids) [54]. Heterogeneous dislocation nucleation refers to the generation of dislocations from pre-existing defects. Some examples for such flaws are grain boundaries, precipitates, surface terraces, pre-existing dislocations, pores, or vacancy clusters [177]. Some of those defects such as grain boundaries and pores can act as stress concentrators to locally increase stresses to nucleate dislocations [65], with the potential competition of crack formation from these defects. Others types of defects like vacancy clusters are believed to play a dual role in aiding dislocation nucleation while decreasing the dislocation mobility [49, 178]. More detailed discussions on the competition between dislocation nucleation and crack formation in ceramics can be found elsewhere [51, 179].

As dislocation nucleation requires high stress, circumventing dislocation nucleation has become necessary towards achieving bulk plasticity in ceramics, as most recently demonstrated by mechanically seeded dislocations at room temperature [14] or high temperature [180] or interface design to borrow dislocations from metals into ceramics [181]. These novel approaches use pre-engineered dislocations as sources to promote dislocation multiplication and motion. Yet so far, only one of these approaches has proven to be feasible to scale up to bulk scale for ceramic materials that exhibit good dislocation mobility at room temperature [4]. The details will be further explored in this section.

## 4.2 Dislocation Motion

By applying a shear stress to a dislocation, it will start to move to an adjacent lattice position, given that the stress exceeds the lattice friction stress [13]. The magnitude of this stress depends on various factors such as temperature or pressure, and is called Peierls stress (defined at 0 K when discussing the motion of dislocations) [182]. Under this condition, the motion of a dislocation is viewed as a straight line moving from one energy valley to an adjacent one by overcoming an energy barrier, i.e., the Peierls potential [182]. The height of this barrier is influenced by various factors, but in a first step can be derived from the crystal lattice and the dislocation core structure [13]. Therefore, the next subsection will explore the different dislocation core structures in several representative room-temperature deformable ceramics. After that, dislocation behavior at finite temperatures will be discussed based on the kink-pair mechanism.

### 4.2.1 Dislocation Core Structure

The dislocation core structure describes the atomic arrangements around dislocations, as atoms are displaced from their usual lattice positions. Analyzing the dislocation core structure requires techniques with atomic resolution. Experimentally, this requires high-resolution transmission electron microscopy (HRTEM) [183, 184] , high-angle annular dark-field scanning transmission electron microscopy (HAADF-STEM) [185] or annular bright-field scanning transmission electron microscopy (ABF-STEM) [185]. Those techniques are able to capture the local atomic structure information about dislocation cores, which are crucial for understanding the mechanisms of dislocation motion and multiplication. In simulation, the core structure was analyzed by calculating the generalized stacking fault (GSF) energies of the slip planes, then using the results as input in the Peierls-Nabarro model, which yields insights into the spatial distribution of the dislocation core [122, 186, 187]. As this only considered atomic displacement in the slip plane, later the Peierls-Nabarro-Galerkin model was employed to obtain information about out-of-plane displacement of atoms around a dislocation. Most recent works are utilizing atomic scale simulation to obtain relaxed models of dislocation cores. These models can also be used to analyze the stresses required for dislocation motion [136, 138, 160, 188].

In most textbooks, the first image of a dislocation appears as edge type, residing at the end of an *as if* inserted extra half plane in the lattice [13, 182]. For some materials like bcc metals this may hold true. However, a full inserted half plane exerts a large elastic distortion in a very small volume, and in fcc metals, such a compact dislocation will dissociate into two partial dislocations, distributing the lattice distortion while forming a stacking fault between them [182], see **Fig. 4B** for simulation in $SrTiO_3$ [138]. This not only reduces lattice distortion but usually allows for easier glide of a dislocation in its glide plane [13]. The distance of partial dislocations is however given by the glide plane's stacking fault energy, as the elastic energy saved from dissociation is in equilibrium with the energy required to form a stacking fault [13]. This is where many ceramics differ from metallic materials, as the stacking fault energies can be much larger, due to electrostatic interactions between the ions [126]. Past efforts in the visualization of dislocation cores in ceramic crystals nicely demonstrated the differences in the core morphology between different crystal structures [189]. In the following, we summarize the insights of both experiments and simulation on the dislocation core structures of the previously presented ceramic compounds.

For ceramics with the fluorite structure, due to the very high stacking fault energy, dislocations exhibit a compact core with a width of roughly one Burgers vector [184, 190]. Unfortunately, the dislocation cores in $CaF_2$ and $BaF_2$ have not been an object of research in recent years, so the exact configuration of the atoms in the core regions remains elusive. A good comparison, however, are the dislocation cores of $UO_2$ with fluorite structure, although $UO_2$ does not plastically deformable at room temperature at macroscale. Atomic simulations by Borde et al. [160] have recently showcased the intricate nature of the dislocation core, consisting of alternating left- and right-oriented U-O bonds. Whether the dislocation core in $CaF_2$ and $BaF_2$ are indeed comparable remains to be validated. This precise analysis of the core structure does highlight the complexity of atomic arrangement in the "simple" compact core, even in binary compounds.

In most rock salt structure crystals, the core structure complexity is further demonstrated, as their dislocation core is neither compact nor dissociated, but extended [115, 121]. This means that while the core spreads out over roughly several unit cells, it does not form a stacking fault in that area (**Fig. 4A**) [188]. While this area looks rather distorted, it forms no stacking fault due to the electrostatic forces acting between the ions, as has been demonstrated by various GSF calculations [121, 122]. This holds true for most rock salt structure crystals, with the exception of AgCl. In this compound, the {110}<110> dislocations split into partials and form a stacking fault in between them [115]. Furthermore, the {001}<110> and {111}<110> slip systems are active at room temperature (**Table 2**). The reason behind this is proposed to be the more covalent nature of the Ag-Cl bond, which sheds some insights into the influence of the bonding conditions on the mechanical material behavior [115, 191]. The extended core structure of two edge dislocations in MgO, captured using TEM is displayed in **Fig. 5A**.

In the perovskite oxide $SrTiO_3$, the {110}<110> dislocations split into partials as well, producing a stacking fault in between them, as depicted in **Fig. 4B** [138]. As the stacking fault energy is still relatively large [192, 193], the partial dislocations are separated by merely several unit cells, and have been reported to be able to move individually to some extent under the right conditions [131, 136, 194]. The exact core structure and its electrostatic nature, being charged or neutral, of the mobile edge dislocations in $SrTiO_3$ has been under debate [136], with a recent study suggesting the dominating charge-neutral nature in atomistic simulation [138]. It was also indicated that a compact core in $SrTiO_3$ would not be mobile [138], underlining the beneficial effect of dissociation on the mobility of dislocations. Furthermore, the dissociated dislocation core (**Fig. 5B**) was imaged previously by the current authors using TEM [19].

The dramatic effect of core reconfiguration on the dislocation mobility is further demonstrated by the photoplastic effect in sphalerite crystals [195]. In the sphalerite structure, the stacking fault energy is low, allowing for larger dissociation distances between partials, with reports demonstrating the individual motion of the partial dislocations [150, 196]. Furthermore, the partial dislocations, depending on their type, may be partially charged, allowing for interactions with charge carriers [87, 195]. In darkness, where there is an absence of photo-excited charge carriers, dislocations are fairly mobile, contributing to large macroscopic plastic strains. The core, as depicted exemplary for the 60° Zn-core in ZnS in **Fig. 4C**, undergoes reconstruction upon interaction with charge carriers as depicted in **Fig. 4D**, which increases the Peierls potential and reduces the dislocation mobility [137]. The intrinsic 60° Zn-core, captured with ABF-STEM is displayed in **Fig. 5C** [137]. As a comparable photoplastic effect is observed for the materials in the wurtzite structure such as ZnO, a similar core reconstruction upon light illumination is assumed [152].

So far, not much has been reported about the room-temperature dislocation core structure in the cesium chloride or the spinel structure. It was argued that the dislocations in the cesium chloride structure do not dissociate but form a compact core instead, owing to their observed glide direction [163]. An analysis using TEM or simulation is however yet missing. This is similar to the dislocation behavior of the spinel structure, where the room-temperature active slip systems are not determined completely, so are the dislocation core structures. It was proposed that the dislocation may dissociate into four partials with three stacking faults connecting them [197]. While this has never been directly observed, it poses many open questions for the dislocations in this structure.

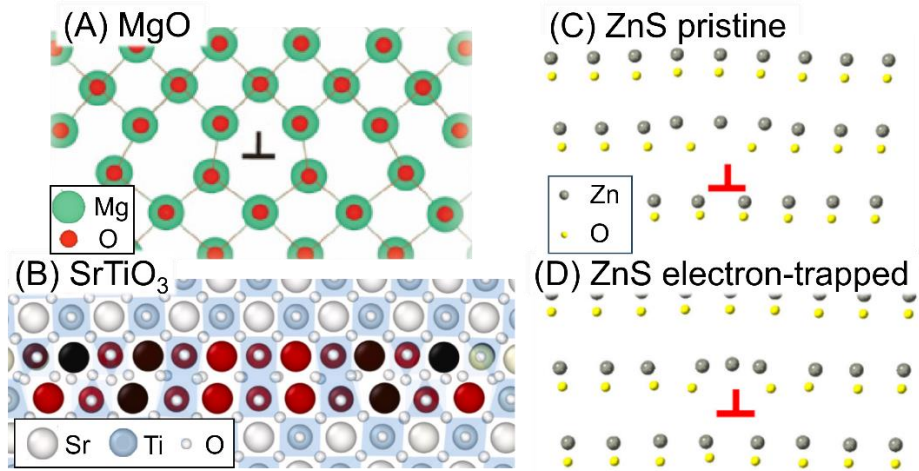

**Figure 4.** (A) Extended core of an {110}<110> edge dislocation in the rock salt structure (MgO). [188] (B) Partial dislocations and stacking fault of a {110}<110> edge dislocation in the perovskite structure (SrTiO$_3$) [138] (C) Core structure of a {111}<110> Zn-core dislocation before and (D) after electron trapping [137]. Images reprinted and modified with permission from (A) [188], (B) [138], and (C,D) [137].

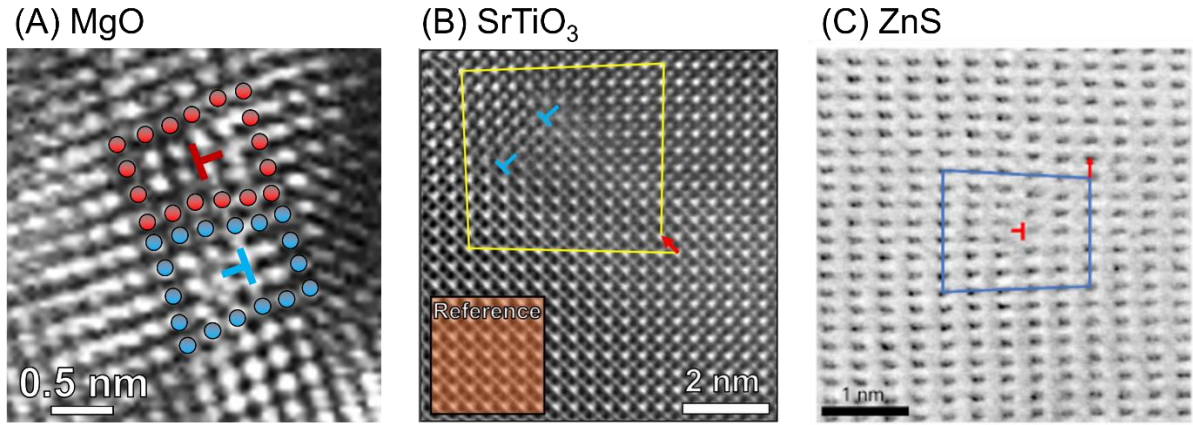

**Figure 5.** (A) Transmission electron microscopy image of two edge dislocations in MgO with atomic resolution. (B) Dislocation core in SrTiO$_3$ exhibiting the dissociated partial dislocations [14]. (C) ABF-STEM image of the intrinsic dislocation core in ZnS [137]. Images reprinted and modified with permission from (B) [14] and (C) [137].

After examining the various edge dislocation core structures of deformable ceramics with different crystal structures, some insights can be gained: while the crystallographic structure is mostly responsible for the possible active slip systems, the mobility of a dislocation is most closely dependent on the atomic arrangement at the dislocation core, which is governed by the atomic species of the compound and the covalency or ionicity of their bonds. Understanding the motion of a dislocation in those compound crystals therefore requires knowledge about the atomic structure and local bonding environment, as those factors influence the height of the Peierls barrier. Yet the complexity brought by the possible charge features of dislocations in ionic/covalent crystals, which inevitably will influence the local bonding environment, requires also careful consideration in the future. Furthermore, the existing literature has mostly focused on the idealized edge dislocations. While they are crucial in the deformation, dislocations of screw or mixed type [14] need to be considered as well for a complete analysis of the deformation behavior.

### 4.2.2 Kink-pair mechanism

After discussing the core structure as a factor influencing the Peierls barrier, it is then of interest to analyze how a dislocation overcomes such a barrier. As known from bcc metals, the kink-pair mechanism plays an important role in overcoming the Peierls barrier [13] and has been demonstrated to lower the critical resolved shear stress by up to two orders of magnitude in ceramic materials as well [98, 198, 199]. The kink-pair mechanism makes dislocation motion via kink formation and migration. The advancement of a short dislocation segment to the next atomic position whilst forming a kink pair requires much lower energy. The sideways motion of the kink pairs to propagate the rest of the dislocation line is energetically more favorable [13, 182, 200]. This is depicted in **Fig. 6A**, where a dislocation line overcomes the Peierls barrier by kink formation into the next Peierls valley [201]. The activation of the kink-pair mechanism has been illustrated to explain the decrease in critical resolved shear stress in most rock salt [98, 122] and perovskite crystals [72, 202]. Furthermore, kink formation and propagation were indicated to influence the glide properties in the fluorite structure and has been investigated with atomistic simulation [160]. Lastly, the kink-pair mechanism has also been investigated in the sphalerite structure, as directly observed in ZnS via in-situ TEM, although ZnS has partial dislocations that are kinking individually [87].

For a better understanding of the dislocation glide behavior at low temperature, it is necessary to underline the kink-pair mechanism and its activation [13], as the kink formation and propagation are thermally activated processes. Information about the kink-pair nucleation energy can help determine whether the dislocation motion is limited by kink formation or by kink propagation, which in turn can be of interest in modeling the velocity of dislocations [200]. Experimentally, this kink-pair nucleation energy can be assessed by means of mechanical spectroscopy through analysis of the Bordoni peaks [203-205]. Theoretically, this energy can be addressed through various models, for example the line tension model [13] or the elastic interaction model [206], which have been employed to model the kink formation of some of the previously presented crystal structures [122, 202]. Furthermore, atomistic simulation has also been used to assess the kink formation and propagation energy in metals and semiconductors [207, 208]. With atomistic simulation, more information about the kink-pair mechanism in ceramics could be achieved, which may further demonstrate its role in the room-temperature deformability of the presented ceramic compounds.

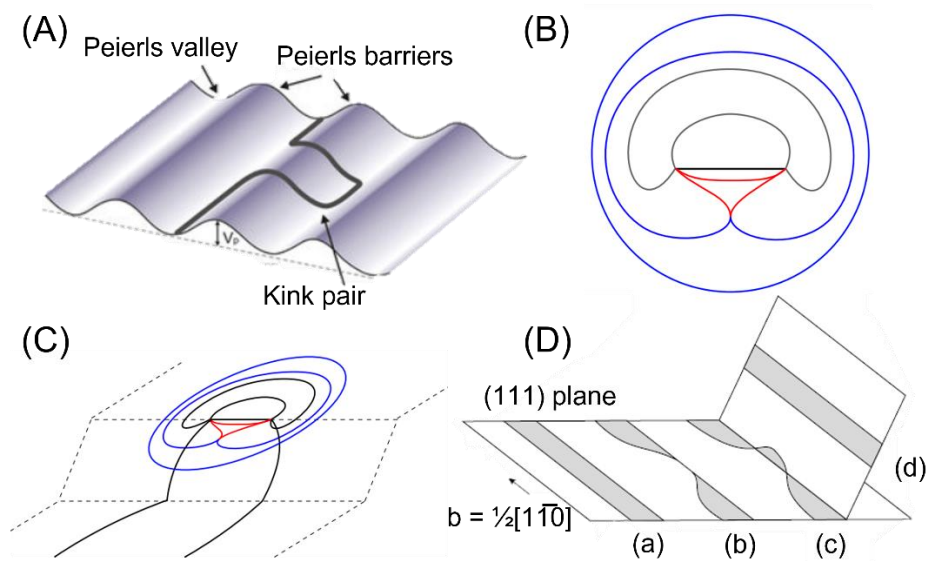

**Figure 6.** Schematic depictions of the important dislocation mechanism in ceramics. (A) A Kink-pair overcoming the Peierls barrier into the adjacent valley [201]. (B) Different stages of multiplication through a Frank-Read source from black over green to read and blue [209]. (C) Multiplication as a result of a cross-slipped dislocation section [209]. (D) Cross-slip of a dissociated dislocation according to the Friedel-Escaig mechanism [182]. Images reprinted and modified with permission from (A) [201], (B,C) [209], and (D) [182].

perovskite crystals [72, 202]. Furthermore, kink formation and propagation were indicated to influence the glide properties in the fluorite structure and has been investigated with atomistic simulation [160]. Lastly, the kink-pair mechanism has also been investigated in the sphalerite structure, as directly observed in ZnS via in-situ TEM, although ZnS has partial dislocations that are kinking individually [87].

For a better understanding of the dislocation glide behavior at low temperature, it is necessary to underline the kink-pair mechanism and its activation [13], as the kink formation and propagation are thermally activated processes. Information about the kink-pair nucleation energy can help determine whether the dislocation motion is limited by kink formation or by kink propagation, which in turn can be of interest in modeling the velocity of dislocations [200]. Experimentally, this kink-pair nucleation energy can be assessed by means of mechanical spectroscopy through analysis of the Bordoni peaks [203-205]. Theoretically, this energy can be addressed through various models, for example the line tension model [13] or the elastic interaction model [206], which have been employed to model the kink formation of some of the previously presented crystal structures [122, 202]. Furthermore, atomistic simulation has also been used to assess the kink formation and propagation energy in metals and semiconductors [207, 208]. With atomistic simulation, more information about the kink-pair mechanism in ceramics could be achieved, which may further demonstrate its role in the room-temperature deformability of the presented ceramic compounds.

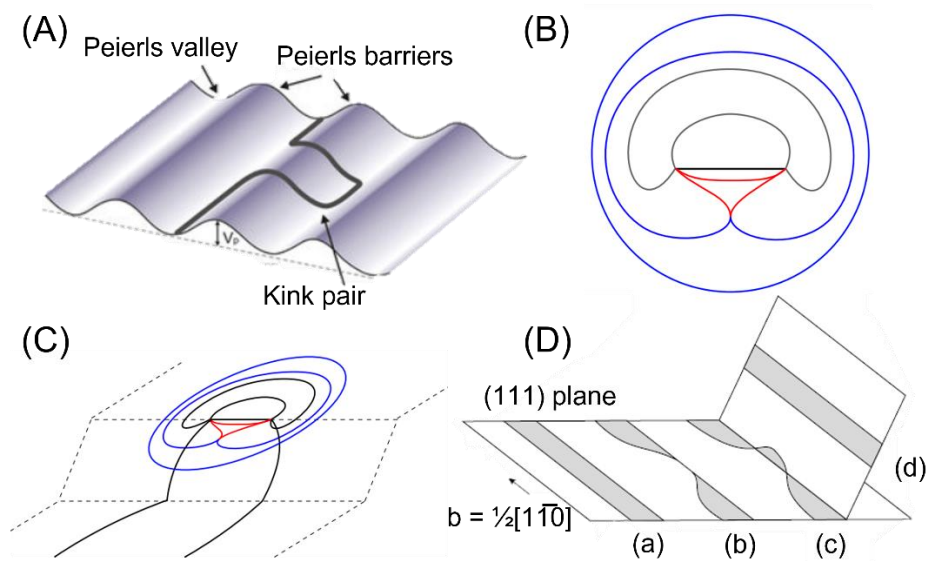

**Figure 6.** Schematic depictions of the important dislocation mechanism in ceramics. (A) A Kink-pair overcoming the Peierls barrier into the adjacent valley [201]. (B) Different stages of multiplication through a Frank-Read source from black over green to read and blue [209]. (C) Multiplication as a result of a cross-slipped dislocation section [209]. (D) Cross-slip of a dissociated dislocation according to the Friedel-Escaig mechanism [182]. Images reprinted and modified with permission from (A) [201], (B,C) [209], and (D) [182].

## 4.3 Multiplication

As dislocation nucleation has been demonstrated as not favorable for bulk-scale introduction of dislocations into ceramics [14], in addition to good dislocation mobility, dislocation multiplication is critical for achieving large plasticity. Dislocation multiplication describes mechanisms that increase the overall total line length in a given volume, hence an increase in the dislocation density. For room-temperature deformation, the Frank-Read mechanism and the cross-slip mechanism are believed to be the dominant mechanism.

A Frank-Read dislocation source consists of a dislocation segment that is pinned at two points, as depicted in **Fig.6B** [209]. Upon applying a shear stress to the pinned segment, depicted in black, it will bow out, loop around the two pinning points and create two parallel sections, which will annihilate upon meeting each other to create a dislocation loop, as depicted in blue and the original pinned section in red. This process is self-repetitive and effective in increasing the dislocation density. Many new dislocation loops can be created on the same slip plane [209, 210]. The pinning points can have different origins, for example, dislocation nodes or junctions, or even precipitates. When the origin is a cross-slipped dislocation segment, it can lead to the deformation of parallel slip planes and activate the multiple cross-slip mechanism [123].

As depicted in **Fig. 6C** [209], the multiple cross-slip mechanism leads to dislocation multiplication on parallel slip planes [123]. This mechanism is important to achieve plasticity throughout a crystal, as it leads to increased dislocation density on other slip planes than the one with the original dislocation source. Unfortunately, cross-slip, which is necessary for the multiple cross-slip mechanism, does not appear to be easily observable in some of the room-temperature deformable ceramics. For instance, uniaxial bulk compression of all three perovskite oxides (**Table 3**) form slip bands within which high-density dislocations are formed, while the adjacent regions remain dislocation-free. This leads to the formation of discrete and inhomogeneous dislocation distribution (see **Fig. 2E**) [15, 71, 86]. This may be an indication of hindered cross-glide behavior and severely limits homogeneous dislocation engineering in bulk-scale samples. In bcc metals, except at very low temperatures, dislocation glide is often non-crystallographic with a wavy glide behavior, which is attributed to cross-slip events and serves as macroscopic evidence for active cross-slip behavior [211]. For ceramic materials, such behavior has only been reported in silver halides and ceramics with the cesium chloride structure [212-214].

Reasons for the less common cross-slip behavior in $SrTiO_3$, besides the limited amount of active slip systems, might be the dislocation dissociation. The dissociation into Shockley partials aids in the motion of the dislocation in the slip plane, due to the smaller elastic distortion of the lattice compared to the perfect dislocation. However, it can also inhibit the cross-slip behavior of screw dislocations, as dislocations would need to recombine according to the Friedel-Escaig mechanism to leave the plane it was gliding on [13, 182]. This is schematically depicted in **Fig. 6D** [182], where it is displayed how a dissociated dislocation in (a) has to constrict to recombine in (b), so that the recombined section can cross-slip in (c) for the dislocation to finally continue gliding on another slip plane in (d) [182]. As recently demonstrated in simulation [138], recombining a dissociated dislocation in $SrTiO_3$ requires a high energy (78 meV/Å) and would severely hinder the mobility of dislocations. Since easier dislocation motion is preferred for plastically deforming ceramics, inhibited cross-slip may obstruct the multiple cross-slip mechanism for dislocation multiplication. This not only influences the dislocation structures after deformation, but also the dislocation density.

Here, we demonstrate two techniques for probing the dislocation density and dislocation structures, as a result of effective dislocation multiplication at room temperature. One example is the classic dislocation etch pit method, in which a chemical etchant specifically attacks the sample surface, where dislocation lines terminate. **Fig. 7(A-C)** show scanning electron microscopy (SEM) images of dislocation etch pits on a $SrTiO_3$ sample after cyclic Brinell indentation. With increasing number of cycles, the etch pit density increases, which translates to an increasing dislocation density of up to ~$10^{13}$/$m^2$ [16]. Dislocation lines can also be seen in transmission electron microscopy (TEM) lamellae, as illustrated in **Fig. 7(D-F)** as the white line contrasts, featuring the increase in dislocation density up to ~$10^{14}$/$m^2$ with increasing number of Brinell scratching passes [215].

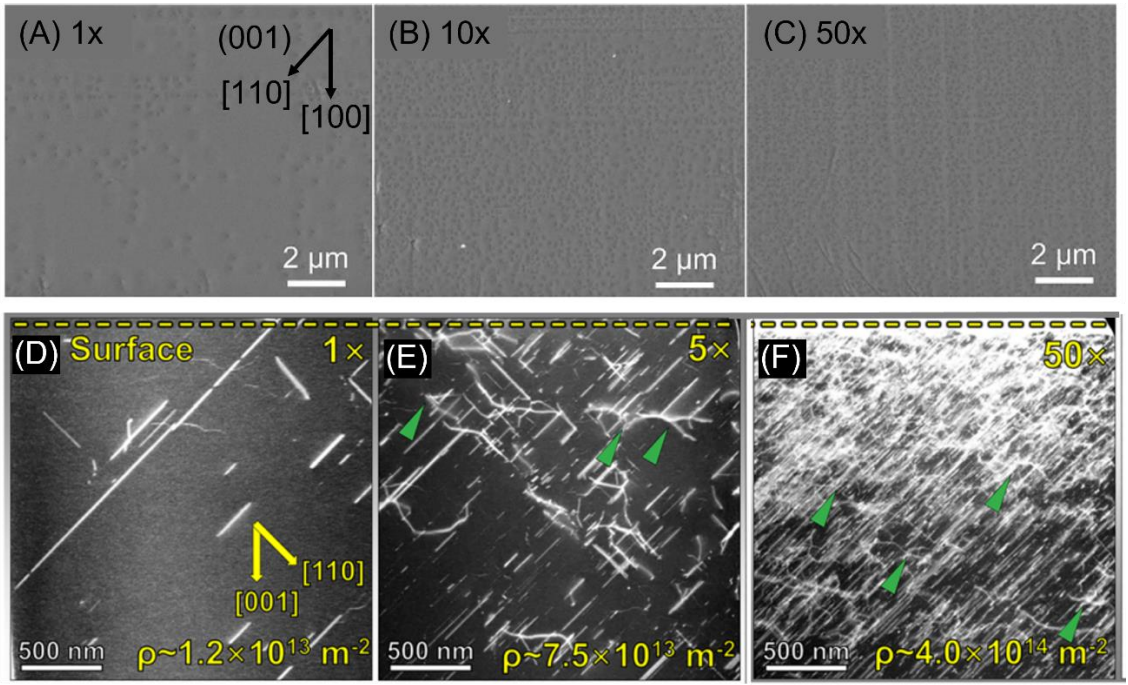

**Figure 7.** (A-C) Scanning electron microscopy (SEM) images of dislocation etch pits from the center of cyclic Brinell indentation on a $SrTiO_3$ (001) single crystal surface with (A) 1 cycle, (B) 10 cycles, and (C) 50 cycles. (D-F) Transmission electron microscopy (TEM) images of lamella taken from Brinell scratch tracks on a $SrTiO_3$ (001) single crystal surface after (D) 1 pass, (E) 5 passes, and (F) 50 passes. The green triangles point towards dislocation jogs. Images reprinted and modified with permission from (A-C) [16], (D-F) [215].

## 5. Discussion

The prediction of dislocation-mediated plastic deformation in ceramics at room temperature has not been properly addressed so far and requires descriptors by which these ceramics can be categorized. In this review, we started with the crystal structure for categorization, which turned out to be hardly a useful descriptor for prediction, as there are many examples of ceramics that have the listed crystal structures but exhibit no dislocation activity at ambient conditions, for example rock salt structure TiC or fluorite structure $UO_2$. A core feature to be examined is the dislocation core structure, which is believed to play a central role for regulating the versatile physical and mechanical properties [216]. Despite the tremendous amount of effort, little is known about the dislocation cores in most of the plastically deformable ceramics at room temperature due to the challenges in imaging them with atomic resolution. As dissociation of dislocations facilitates in-plane dislocation motion, as seen from the studies on $SrTiO_3$, the ability to form partial dislocations appears to provide insights into the good mobility. However, it was shown for $BaTiO_3$, which is believed to exhibit bulk dislocation plasticity only at high temperature, that dislocations may also dissociate [186], adding uncertainties of using solely the dislocation core structure as an indicator for room-temperature plastic behavior.

For metals, it was stated by Pugh in 1954 that the degree of plastic deformability could be estimated by comparing the ratio of bulk modulus $K$ to shear modulus $G$ [217]. The larger $K/G$, the more deformable a metal is predicted to be, which is still being used to predict the deformability of various materials [218-220]. While this proved useful to a certain extent for sufficiently isotropic, polycrystalline metals with multiple active slip systems, this approximation fails for the highly anisotropic, single-crystalline ceramics, which have fewer active slip systems at room temperature. For example, TiC yields a $K/G$ ratio of 1.36 [221], comparable to metallic Zn with 1.59 [217] and lower than the readily deformable MgO of the same structure with 1.31 [222], without displaying any sign of dislocation-mediated plasticity at room temperature. It remains therefore questionable, how well ceramics with a large $K/G$ ratio can actually deform plastically. Similar models exist for the ratio of theoretical strength to theoretical shear strength [223, 224], or based on the velocity of sound [225], however with similar outcome. As was stated by Thompson and Clegg in 2018: "no simple model that works on the basis of simple relations of bulk polycrystalline properties can represent the failure mode of different materials" [226], adding to the doubt about using the $K/G$ ratio as a reliable measure for ductility prediction. They furthermore state the necessity to consider the dislocation itself in the particular structures to adequately predict deformability, which is not an easy endeavor [226].

In the particular case of deformable ceramics, another clue may be the covalency of the bond. As demonstrated by Nakamura et al. [115], AgCl is more easily deformable than NaCl, due to the larger number of active slip systems, which has been attributed to the more covalent nature of the bond, reducing the repulsive electrostatic forces from bringing likewise charges close to each other during glide. Purely covalent bonding, however, can also not be the descriptor of plastic behavior, as diamond, bonded purely covalently, is the hardest material known in nature. Here, the strong covalent bonds would need to be broken and reconnected each time a dislocation progresses. Therefore, a partly covalent, partly ionic bonding situation seems to be helpful in the motion of dislocations in ceramic materials, as further demonstrated by dislocation glide in perovskite oxides $ABO_3$. Here the A-O bond is of mostly ionic character and the B-O bond is a mixed ionic-covalent bond [227]. As seen in the most recent analysis of the dislocation core in $SrTiO_3$ [138], from the depictions, it appears that for dislocation motion, the B-O bond is broken and reconnected, indicating the influence of the partly covalent bond on the glide of dislocations. Nevertheless, open questions remain for this approach. If the dislocation motion is only dependent on the partly covalent B-O bond, then why does changing the B-site from Nb to Ta in $KNbO_3$ not suppress the deformability? Why does changing the A-site from K to Na reduce the deformability, resulting in the not deformable compound $NaNbO_3$ in bulk? Most recently, the bonding environment of various intermetallic compounds has been analyzed and successfully correlated to their plastic properties at small scales [228]. It was proposed that for a further understanding the plastic deformation behavior as well as predictions about room-temperature ductility in ceramics, an atomistic approach focusing on the bonding conditions at the dislocation core is necessary.

Last but not least, we briefly address the competition between plasticity and crack formation in ceramics. For crack formation, it can be helpful to separately examine crack initiation and crack propagation. Regarding crack initiation, it has been experimentally observed that dislocation-dislocation interaction [229, 230], dislocation-kink bands interaction [231], and dislocation pileup at grain boundaries/interfaces [179] (namely, the Zener-Stroh model [232-234]) can lead to crack initiation. Therefore, it can be expected that achieving ultra-high dislocation densities may continue be to a big challenge for effective crack suppression.

Concerning crack propagation in ceramics, with a crack tip subjected to increasing stress intensity factor, the pertinent question that had been long raised was whether the crack will directly propagate, or the crack tip will be blunted by spontaneous dislocation emission prior to crack propagation. The most classical analytical model to describe this scenario is the Rice-Thomson model [12], which was later modified by Rice to include the Peierls potential [235]. In essence, the model compares the energy needed to nucleate a single dislocation from the crack tip (which is given as a function of the stacking fault energy) with the fracture energy (which is a function of the specific surface energy) [236]. As measuring these two properties is experimentally challenging, literature values for both can be off by an order of magnitude, making the method only applicable for extreme cases. Moreover, after the first dislocation is emitted, its stress field will influence the second one, increasing the necessary emission stress with every new dislocation. Other pre-existing surrounding defects can also hinder the dislocation motion, as described above. This is not accounted for in the Rice model, which considers the very first emitted dislocation in a perfect crystal.

In short, unlike in most metals, so far no effective crack tip blunting, even with spontaneous dislocation emission possible, has been found in ceramics such as in MgO and NaCl [237, 238] that exhibit excellent room-temperature dislocation plasticity. The dislocation shielding model developed by Higashida et al. [237, 238] for NaCl and MgO yields marginal fracture toughness increase by dislocations. The competition between dislocation plasticity and cracking initiation as well as crack tip – dislocation interaction in ceramics requires a more detailed mechanistic understanding to predict whether a ceramic sample will plastically deform or fracture under given mechanical loading.

## 6. Summary & Outlook

To summarize, the most relevant mechanical deformation methods across the length scales and 44 ceramic materials are presented for their room-temperature bulk dislocation plasticity. The bulk plasticity suggests a low lattice resistance for the dislocations to glide at room temperature. The critical role of dislocation core on the dislocation mechanisms including nucleation, motion, and multiplication is examined, and open questions for the origin of dislocation plasticity in ceramic materials at room temperature are posed. As most recently showcased on bulk compression of $KTaO_3$ as the third plastically deformable perovskite oxide at room temperature [19], there can be more of such ceramics to be discovered. Not only does the analysis of new deformable ceramics help understand the plasticity of ionic/covalent crystals, it will also uncover new opportunities for the proposed dislocation technology in ceramics. Achieving processing and deformation at intermediate/room temperature will significantly reduce the energy and time cost, rendering it a worthwhile endeavor for discovering more *ductile* ceramics at ambient temperatures. The combinatorial experimental deformation toolbox summarized here can yield effective testing for ceramics, which is now being extended to coarse-grained polycrystalline oxides [65]. Nevertheless, the current approaches reviewed here are still of a trial-and-error nature. With the help of many new ceramics, especially perovskite oxides, being discovered [239], experimental data can now be more efficiently collected to lay the groundwork for predicting room-temperature dislocation plasticity in ceramics at the macroscale and for constructing the desired materials toolbox for ductile ceramics.

**Acknowledgement:**

Funding from the European Union (ERC Starting Grant, Project MECERDIS, grant No. 101076167) and is acknowledged. Views and opinions expressed are, however, those of the authors only and do not necessarily reflect those of the European Union or the European Research Council (ERC). Neither the European Union nor the granting authority can be held responsible for them. Deutsche Forschungsgemeinschaft (DFG, grant No. 510801687 and 414179371) is also acknowledge for financial support. W. Lu acknowledge the support by Shenzhen Science and Technology Program (grant no. JCYJ20230807093416034).

**Conflict of interest statement:**

The co-authors, Atsutomo Nakamura and Xufei Fang, are serving as guest editors for the special issue "*Dislocations in Ceramics*", to which this manuscript was invited by the Editor-in-Chief. The handling editor and reviewers are assigned independently by the Editor-in-Chief to avoid potential conflict of interest.